# Terahertz Digital Reconfigurable Metamaterial Array for Dynamic Beamforming Applications

Eistiak Ahamed[#1], Rasool Keshavarz[#2], Negin Shariati[#3]

[#]RF and Communication Technologies (RFCT) Research Laboratory, School of Electrical and Data Engineering, Faculty of Engineering and Information Technology, University of Technology Sydney, Broadway, Ultimo, NSW, 2007, Australia.

[1]eistiak.ahamed@student.uts.edu.au,

{[2]rasool.keshavarz, [3]negin.shariati}@uts.edu.au

*Abstract* — A novel digital reconfigurable 2-bit metamaterial, equipped with a substrate integrated feeding system, is designed for industrial quality control applications within the terahertz frequency range. The proposed feeding mechanism facilitates azimuthal beam steering, spanning from −90° to +90°, thereby enabling the reconfiguration of beam patterns within the digital metamaterial. Utilizing the phase distribution concept and comprehensive analysis of coupling and the e-field effect on individual unit cells, the metamaterial array spacing is meticulously designed. Operational at 0.7 THz, the system offers versatile reconfigurability, supporting single, dual, and multibeam modes. Through meticulous optimization, the system demonstrates an impressive −138° to +138° beam steering capability. This dynamic beamforming ability, transitioning seamlessly from a singular beam to multibeam configurations, requires minimal software-hardware integration for scanning and inter-satellite links, thus presenting significant potential for enhancing product quality within industrial environments and satellite communication.

## I. Introduction

The utilization of terahertz (THz) beamforming has garnered considerable attention in recent years among researchers seeking to augment the speed and intelligence of advanced devices. Within the domain of sophisticated wireless communication systems, inter-satellite links and scanning leveraging THz technology [1, 2] emerge as a promising avenue to address pressing challenges, including spectrum scarcity, misalignment fading, path loss, and intricate security concerns, within the dynamic landscape of evolving 6G networks[3, 4]. The promise of THz lies in its vast spectral resources available in the THz band, offering high-penetration propagation capabilities across a broad spectrum ranging from 0.1 to 10 THz [5]. Nevertheless, unlocking this potential necessitates a profound grasp of the intricate technical intricacies, complex regulatory frameworks, and ongoing interaction of emerging innovations at the forefront of wireless and THz technology research.

Modern beamforming devices, constructed from traditional unstructured bulk materials, frequently encounter challenges like notable losses, restricted sensitivity, bandwidth, and dynamic capabilities[6]. Metamaterials, with their inventive artificial periodic structures comprising metal-dielectric combinations, present a versatile and economical alternative to address these shortcomings [7-11]. Engineered 1-bit arbitrary coded metamaterial was designed for wideband functionality spanning 0.8 to 1.4 THz. Through the application of a refined methodology, this metamaterial achieves precise control over beam steering, intricately minimizing reflection within a designated angular range of 0° to 40° across multiple frequency bands [12]. At the forefront of technological innovation, a ground breaking metamaterial has been introduced, designed to dynamically manipulate beamforming processes at the remarkable frequency of 0.3 THz [13]. Offering a versatile functionality, this metamaterial facilitates both linearly and circularly polarized metasurfaces, promising transformative applications in the realm of Tera-WiFi links. Leveraging advanced principles of electromagnetic manipulation, the beam can be precisely steered across a wide angular spectrum, ranging from −45° to +45°, thereby paving the way for unprecedented control and optimization in terahertz communication networks. Furthermore, by harnessing the intricate properties of metamaterials, this pioneering approach holds the potential to significantly enhance the efficiency of THz systems, revolutionizing the landscape of high-speed wireless communication through the ingenious utilization of metamaterial-aided surfaces[14].

In this paper, a novel 2-bit digital metamaterial is employed alongside a substrate-integrated log periodic antenna to construct a dynamic digital metamaterial beamformer. The proposed reconfigurable metamaterial array features innovative digital reconfigurability, flexible beam steering capabilities, GaAs THz Field-Effect Transistors integration, and potential for industrial quality control and improved inter-satellite link for satellite network applications, establishing it as a state-of-the-art technology in terahertz communication networks. Through intricate design and synchronization strategies, this metamaterial beamformer showcases a dynamic beamforming pattern tailored to accommodate a range of bit sequences, highlighting its adaptability in diverse scenarios. The integrated feeding design is meticulously crafted to minimize coupling effects and ensure precise phase differences within the digital metamaterial arrays, resulting in a comprehensive system housed within a single substrate, poised to address industrial-quality applications with unparalleled efficiency and precision. Furthermore, the developed system demonstrates robust performance in industrial environments, where stringent quality standards are paramount. By integrating advanced technologies within a single substrate, such as the 2-bit digital metamaterial and log periodic antenna, this innovative approach promises streamlined implementation and enhanced scalability for sensing, image recontraction, cryptography, information encoding, inter-satellite link, and smart agriculture applications.

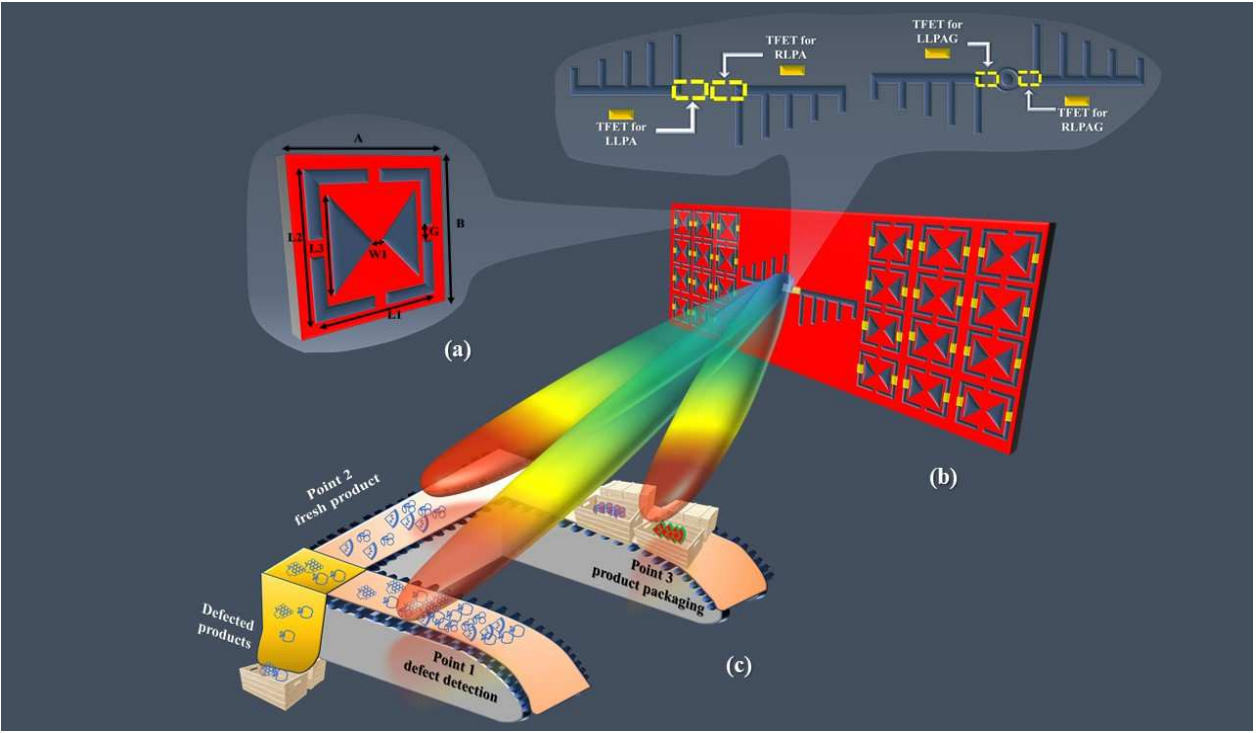

Fig. 1. (a) Proposed metamaterial unit cell design where A is 130 $\mu m^2$, B is 130 $\mu m^2$, L1 is 130 $\mu m^2$, L2 is 120 $\mu m^2$, L3 is 80 $\mu m^2$, W1 is 10 $\mu m^2$ and G is 8 $\mu m^2$, (b) the proposed beamforming system where inset the log preodic antenna mechanism is zoomed RLPA and LLPA is the top radiator for antenna and RLPAG and LLPAG indicated the ground of the antenna where yellow strip is indicated the GaAsTFET, and (c) the application scenario for industrial quality control at THz frequency range.

## II. WORKING CONCEPT

The proposed reconfigurable system relies on digital reconfigurable metamaterial to facilitate beam steering technology, allowing for variable angles across different sequences. Designed as the source to feed the digital metamaterial, the Log Periodic Antenna (LPA) plays a crucial role in enabling this functionality. The combined utilization of the log periodic antenna and metamaterial array ensures operation within the 6G frequency range at 0.7 THz, illustrating the system adeptness in addressing the demands of contemporary communication paradigms. The LPA is designed based on a proper design guide [15] where the geometry adheres to the following rules

$$\tau = \frac{L_{n+1}}{L_n} \tag{1}$$

$$\sigma = \frac{S_n}{2L_n} \tag{2}$$

$$\alpha = \tan^{-1}\left[\frac{1-\tau}{4\sigma}\right] \tag{3}$$

The $N$ number of periodical active dipoles contained in LPA where $L_n$ is the length of reflector dipole, $\sigma$ is the spacing factor, $\alpha$ is the apex angle, $S_n$ is the spacing factor and $\tau$ is the scaling factor. The field distribution of the log periodic array (LPA) can be defined into three distinct regions, comprising the short arm, active, and reflector regions, each playing a pivotal role in the antenna overall functionality. Estimations concerning the radiation characteristics of the LPA antenna necessitate a thorough analysis of the intricate current distribution patterns observed across the active dipole elements. Furthermore, the designed log periodic antenna exhibits a remarkable feature as a directional beam generator, wherein the electric field emanating from the active dipole region can be precisely assessed at any given distance "r" from its point of radiation [16].

$$E_r = 2E_{max}\left[\frac{\pi R_n\left\{\left(\frac{1}{\tau}\right)-1\right\}}{\lambda}\sin\theta\right] \tag{4}$$

where, $R_n$ denotes the displacement of the feed point perpendicular to the dipole axis.

The Log Periodic Antenna, engineered for operation within the 0.5-1 THz spectrum, serves as the primary feeder for the digital metamaterial unit cell, thus enabling a high degree of reconfigurability. This unit cell, meticulously designed around the phase difference concept [17, 18] and finely tuned to resonate at 0.7 THz in conjunction with the antenna setup, is meticulously tailored for industrial quality control applications, as vividly depicted in Figure 1(c). Moreover, the devised beamformer exhibits remarkable versatility for multibeam applications, offering sophisticated capabilities such as defect detection, real-time segregation of fresh products, and precise scanning for foreign contaminants within product batches. Consequently, this advanced beamforming system not only elevates product safety and quality standards but also streamlines manufacturing processes, while its potential for automation is further enhanced through seamless integration with advanced control units, signal processing modules, and robust feedback loops, ensuring optimal operational efficiency.

### III. DESIGN MECHANISM AND RESULTS

The entire beamforming system is meticulously crafted around two log periodic antennas, designated as the Right-Side Antenna (RA) and the Left-Side Antenna (LA). Within this framework, a novel 3×3 metamaterial array (MA) structure serves as the backbone for beam steering, with an equal distribution of MA units on both the right (MAR) and left (MAL) sides shown in Figure 1(b). Operating on a Silicon substrate and featuring Gold as the metallic component in the metal-dielectric interface, this system embodies a harmonious fusion of advanced materials and engineering principles. The reconfigurable beamforming lens, sized at $1326 \times 624 \ \mu m^2$, and the compact metamaterial lens, occupying a footprint of only $122 \times 120 \ \mu m^2$, epitomize the system's efficient utilization of space and resources. Through the precise activation of GaAs Terahertz Field-Effect Transistors (GaAsTFETs), the proposed metamaterial structure showcases its adaptability and responsiveness to dynamic reconfigurability, positioning it at the forefront of transformative beamforming technologies.

Operating considerations for the system are bifurcated into two categories: one pertaining to the antennas and the other to the metamaterial. Two GaAsTFETs are allocated for each antenna, yielding a 2-bit operating pattern with four sequences. Specifically, "00" signifies both antennas are deactivated, "01" denotes the activation of RA with LA turned off, "10" represents LA activated while RA remains off, and "11" indicates both antennas are active. The same sequence pattern is mirrored for the metamaterial array. To integrate GaAs THz Field-Effect Transistors into a metamaterial structure with precise Split Ring Resonator gaps, advanced techniques like Metal-Organic Chemical Vapor Deposition for material growth, Electron Beam Lithography for patterning, Focused Ion Beam etching, Atomic Layer Deposition for dielectric layers, and high-resolution alignment and nanomanipulation tools can be employed, followed by thorough electrical and terahertz testing.[19]

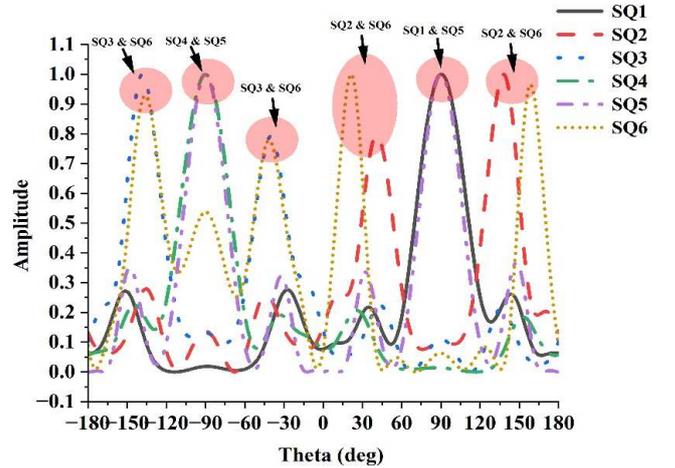

Fig. 3. Beam steering angle for sequences.

The system reconfigurability is precisely outlined through a comprehensive analysis comprising six distinct sequences, meticulously labeled as SQ1 through SQ6. In the complex coordination of these sequences, SQ1 plays a crucial role as the initial step, triggering the activation of RA and MAR, while simultaneously keeping LA and MAL in an inactive state. Transitioning to SQ2, RA is strategically engaged alongside the first column of MAR, while the remaining components of the system are held dormant, showcasing a nuanced approach to beam steering modulation. Similarly, SQ3 governs the activation of LA and the initial column of MAL, exemplifying a meticulous delineation of the operational modalities within the system. Further augmenting the complexity of the system configuration, SQ4 introduces a scenario where LA and MAL are activated, while RA and MAR are purposefully deactivated, thus illustrating the complicated interplay between different components in achieving desired beamforming outcomes. Moving forward, SQ5 introduces a simultaneous activation of both RA and LA, complemented by the engagement of their respective metamaterial inclusions, underscoring the system inherent adaptability to multifaceted operational scenarios. Perhaps most interestingly, SQ6 introduces a configuration where LA is selectively activated alongside the first two columns of MAL, effectively broadening the scope of beam steering capabilities to encompass a wider range of angles compared to SQ2 and SQ3, thereby presenting a more refined multibeam configuration.

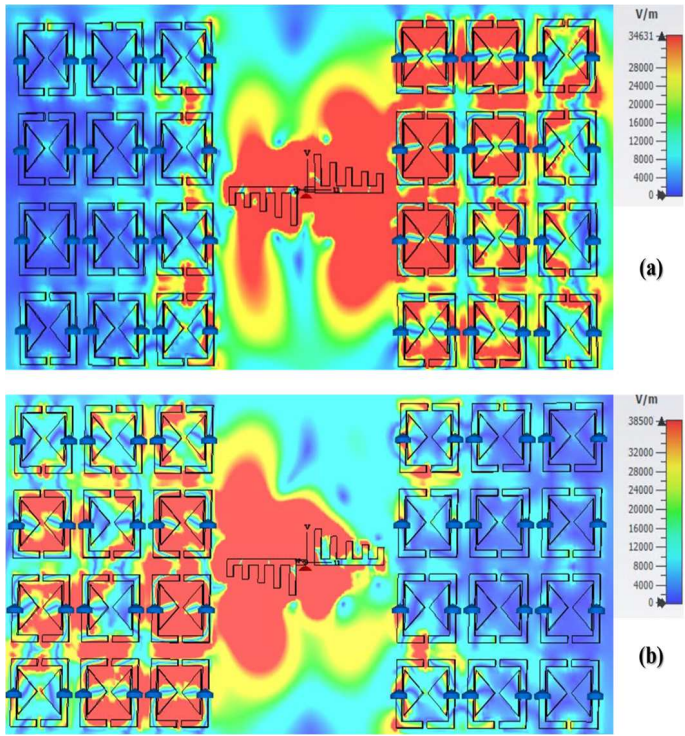

Fig. 2. E-field distribution at 0.7 THz for (a) SQ5, and (b) SQ4

The reconfigurable patterns elucidated within Table 1 provide a comprehensive overview of the system dynamic capabilities, encapsulating the creation of single, dual, and multi-beam patterns across various operational sequences. SQ1 and SQ4, for instance, appropriately demonstrate the realization of single-beam patterns, adeptly navigating within a range from $-90°$ to $+90°$. Conversely, SQ5 showcases the simultaneous generation of beams at $-90°$ and $+90°$, epitomizing the system versatility in adapting to diverse operational scenarios. Delving deeper into the system multifaceted operational modes, SQ2 and SQ3 showcase the generation of dual beams at $+138°$, $+40°$, and $-138°$, $-40°$, respectively, underscoring the system innate capacity for sophisticated beamforming operations. Finally, SQ6 emerges as a testament to the system unparalleled versatility, generating a multibeam pattern that spans a diverse array of angles, including $-136°, -40°, +20°$, and $+158°$, thus elevating the system adaptability to unprecedented levels.

Complementing these elaborate operational modalities, the normalized beam steering pattern elucidated within Figure 3 provides a visually immersive representation of the system performance across different sequences, further enhancing the understanding of its intricate operational dynamics.

Table 1. Reconfigurable responses of digital metamaterial system

| Reconfigurable sequences | Beam steering angle (°) | Gain (dBi) |
|---|---|---|
| SQ1 | $+90$ | 4.77 |
| SQ2 | $+138 \& +40$ | 4.6 & 3.6 |
| SQ3 | $-138 \& -40$ | 5.35 & 4.36 |
| SQ4 | $-90$ | 5.8 |
| SQ5 | $-90 \& +90$ | 2 & 1.2 |
| SQ6 | $-136, -40, +20 \& -158$ | 2.65, 1.9, 2.9 & 2.8 |

IV. CONCLUSION

The tailored 2-bit metamaterial, engineered specifically for industrial quality control scanning applications at the intricate frequency of 0.7 THz, stands out for its distinctive feature of a switchable substrate-integrated feeding mechanism. This innovative system design enhances digital metamaterials by incorporating advanced features, resulting in an extensive beam steering range from $-138°$ to $+138°$. Moreover, its capacity for seamless transition between single-beam and multibeam configurations augments the potential for automated scanning systems and inter-satellite links, underscoring its adaptability to complex operational environments. The strategic integration of a log periodic antenna serves to not only extend the scanning range but also to optimize the systems form factor. With its multifaceted functionalities, ranging from image reconstruction to smart agriculture and airport conveyor systems, this innovative solution representatives a new era in the realm of advanced technological applications.

REFERENCES

[1] I. F. Akyildiz, J. M. Jornet, and C. Han, "TeraNets: Ultra-broadband communication networks in the terahertz band," *IEEE Wireless Communications,* vol. 21, no. 4, pp. 130-135, 2014.
[2] S. Lalithakumari, S. K. Danasegaran, G. Rajalakshmi, R. Pandian, and E. C. Britto, "Analysis of Wave Propagation in Hybrid Metamaterial Structure for Terahertz Applications," *Brazilian Journal of Physics,* vol. 53, no. 5, p. 140, 2023.
[3] U. Nissanov and G. Singh, "Multi-beam and Beamforming Terahertz Array Antenna for 6G Communication," in *Antenna Technology for Terahertz Wireless Communication*: Springer, 2023, pp. 219-262.
[4] K. Tekbiyik, G. K. Kurt, A. R. Ektı, and H. Yanikomeroglu, "Reconfigurable intelligent surfaces empowered THz communication in LEO satellite networks," *IEEE Access,* vol. 10, pp. 121957-121969, 2022.
[5] J. Álvarez-Sanchis, B. Vidal, S. Tretyakov, and A. Díaz-Rubio, "Loss-induced performance limits of all-dielectric metasurfaces for terahertz sensing," *Physical Review Applied,* vol. 19, no. 1, p. 014009, 2023.
[6] C. X. Liu, F. Yang, X. J. Fu, J. W. Wu, L. Zhang, J. Yang, and T. J. Cui, "Programmable manipulations of terahertz beams by transmissive digital coding metasurfaces based on liquid crystals," *Advanced Optical Materials,* vol. 9, no. 22, p. 2100932, 2021.
[7] T. Freialdenhoven, M. Schepers, and T. Dallmann, "Frequency Controlled Polarization Rotating Transmitarray for Polarimetric Radar Applications," in *2022 52nd European Microwave Conference (EuMC)*, 2022: IEEE, pp. 660-663.
[8] M. Lippke, E. Stoja, D. Philipp, S. Konstandin, J. Jenne, T. Bertuch, and M. Günther, "Investigation of a Digitally-Reconfigurable Metasurface for Magnetic Resonance Imaging," in *2022 52nd European Microwave Conference (EuMC)*, 2022: IEEE, pp. 668-671.
[9] M. Ansari, H. Zhu, N. Shariati, and Y. J. Guo, "Compact planar beamforming array with endfire radiating elements for 5G applications," *IEEE Transactions on Antennas and Propagation,* vol. 67, no. 11, pp. 6859-6869, 2019.
[10] R. Keshavarz, N. Shariati, and M.-A. Miri, "Real-Time Discrete Fractional Fourier Transform Using Metamaterial Coupled Lines Network," *IEEE Transactions on Microwave Theory and Techniques,* 2023.
[11] M. Ullah, R. Keshavarz, M. Abolhasan, J. Lipman, and N. Shariati, "Multi-Service Compact Pixelated Stacked Antenna with Different Pixel Shapes for IoT Applications," *IEEE Internet of Things Journal,* 2023.
[12] L. Liang, M. Qi, J. Yang, X. Shen, J. Zhai, W. Xu, B. Jin, W. Liu, Y. Feng, and C. Zhang, "Anomalous terahertz reflection and scattering by flexible and conformal coding metamaterials," *Advanced Optical Materials,* vol. 3, no. 10, pp. 1374-1380, 2015.
[13] W. O. Carvalho, E. Moncada-Villa, J. R. Mejía-Salazar, and D. H. Spadoti, "Dynamic terahertz beamforming based on magnetically switchable hyperbolic materials," *Journal of Physics D: Applied Physics,* vol. 57, no. 17, p. 175001, 2024.
[14] Z. Wang, X. Mu, J. Xu, and Y. Liu, "Simultaneously transmitting and reflecting surface (STARS) for terahertz communications," *IEEE Journal of Selected Topics in Signal Processing,* 2023.
[15] G. H. Koepke, D. A. Hill, and M. T. Ma, "Analysis of an array of log-periodic dipole antennas for generating test fields," 1987.
[16] S. P. KOSTA, "A Theory of the Log-periodic Dipole Antenna," *International Journal Of Electronics,* vol. 23, no. 5, pp. 473-483, 1967.
[17] T. J. Cui, M. Q. Qi, X. Wan, J. Zhao, and Q. Cheng, "Coding metamaterials, digital metamaterials and programmable metamaterials," *Light: science & applications,* vol. 3, no. 10, pp. e218-e218, 2014.
[18] K. Chen, W. Song, Z. Li, Z. Wang, J. Ma, X. Wang, T. Sun, Q. Guo, Y. Shi, and W.-D. Qin, "Chalcogenide phase-change material advances programmable terahertz metamaterials: a non-volatile perspective for reconfigurable intelligent surfaces," *Nanophotonics,* no. 0, 2024.
[19] D. Coquillat, J. Marczewski, P. Kopyt, N. Dyakonova, B. Giffard, and W. Knap, "Improvement of terahertz field effect transistor detectors by substrate thinning and radiation losses reduction," *Optics Express,* vol. 24, no. 1, pp. 272-281, 2016.